\documentclass[psfig]{llncs}
\usepackage{multicol} % needed for the list of participants
\usepackage{makeidx}  % allows for indexgeneration
\usepackage{psfig}
\begin{document}
\frontmatter          % for the preliminaries
\pagestyle{headings}  % switches on printing of running heads
\addtocmark{Jamming Transition} % additional mark in the TOC
\titlerunning{Jamming Transition}
\title{Jamming Transition in CA Models for Traffic Flow}
\titlerunning{Jamming Transition}  % abbreviated title (for running head)
\author{Ludger Santen \and Andreas Schadschneider}
\authorrunning{Ludger Santen et al.}   % abbreviated author list (for running head)
%
%%%% modified list of authors for the TOC (add the affiliations)
\tocauthor{Ludger Santen, Andreas Schadschneider (Universit\"at zu K\"oln)}
\institute{Institut f\"ur Theoretische Physik \\
  Universit\"at zu K\"oln\\ D-50937 K\"oln, Germany}

\maketitle              % typeset the title of the contribution

\begin{abstract}
The cellular automaton model for traffic flow exhibits a jamming
transition from a free-flow phase to a congested phase. In the
deterministic case this transition corresponds to a critical point
with diverging correlation length. We present data from
numerical simulations which suggest the absence of critical behavior
in the presence of noise.
The transition of the deterministic case is smeared out and one
only observes the remnants of the critical point.
\end{abstract}
\section{Introduction}
The cellular automaton (CA) approach to traffic flow theory
\cite{NaSch} has attracted much interest in recent years. Compared to
the earlier attempts in modeling traffic 
flow (see e.g.\ \cite{juelbook} and references therein) CA models can
be used very efficiently for computer simulations. This allows to
perform realtime simulations of large networks \cite{town}.

One of the simplest models of that type is the model introduced by
Nagel and Schreckenberg (NaSch model) \cite{NaSch}. This model is
discrete in space and time. A discrete space means that the road is
divided into $L$ cells which can be occupied by a car or can be empty.

The model contains only a few parameters: The braking noise $p$, the
maximum velocity $v_{max}$ and, if one considers periodic boundary
conditions, the average density $\rho$ of cars. The update is divided into
four steps which are applied in {\em parallel} to all cars. The first
step (R1) is an acceleration step. The velocities $v_j$ of each car
$j=1, \dots, N $ not already propagating with the maximum velocity
$v_{max}$ are increased by one. The second step (R2) is designed to
avoid accidents.  If a car has $d_j$ empty cells in front of it and
its velocity (after step (R1)) exceeds $d_j$ the velocity is reduced
to $d_j$. Up to now the dynamics is completely deterministic. Noise is
introduced via the randomisation step (R3). Here the velocities of
moving cars $(v_j\geq 1)$ are decreased by one with probability $p$.
The steps (R1)--(R3) give the new velocity $v_j$ for each car $j$.  In
the last step of the update procedure the positions $x_j$ of the cars
are shifted by $v_j$ cells (R4) to $x_j+v_j$. 

For a given set of parameters $v_{max}$ and $p$ one can distinguish two
different phases depending on the density
\cite{eisi,Csanyi,Sasvari,Luebeck}. For low densities each 
car can propagate with its desired velocity $v_{max}-p$ such that the
flow is given by $J =\rho(v_{max}-p)$. For higher values of the
average density jams occur and start and stop waves dominate the
dynamics. Here we study the nature of the transition and the behaviour
of the model at the transition point.

\section{Order - Parameter}

For a proper description of the transition one has to introduce an
order-parameter (see also \cite{Vilar}) with a qualitative different
behaviour in the two phases. Due to particle conservation one cannot 
 use the density of
cars, which would be the analogue to the magnetisation in the Ising
model. Among several possible choices of an order-parameter the density
of nearest neighbour pairs
\begin{equation}
  m (\rho) = \frac{1}{L} \sum_{i=1}^{L}<n_in_{i+1}>
\label{order-parameter}
\end{equation}
with  $n_i = 1(0)$  if the site $i$ is occupied (empty), is probably
the simplest choice of a {\em local} order-parameter. One could think
of alternative definitions, e.g. the density of
cars with a distance less than $v_{max}$, but those results are in 
qualitative agreement with results obtained using
(\ref{order-parameter}).
\begin{figure}[h]
 \centerline{\psfig{figure=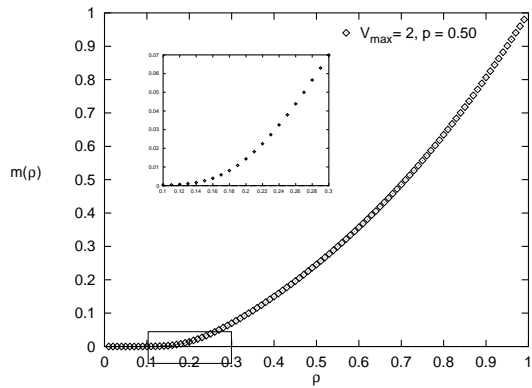,bbllx=95pt,bblly=180pt,bburx=550pt,bbury=510pt,height=5cm}}
\caption{\protect{Behavior of the order-parameter for a finite
      braking probability $p$. It does not vanish exactly for $\rho < \rho_c$,
      but converges smoothly to zero even for small values of the braking 
      probability $p$.}}
\label{fig_op}
\end{figure}
Fig. \ref{fig_op} shows the density dependence of the order-parameter
for $v_{max} = 2$ and $p=0.50$. In the free flow regime the order-parameter
vanishes and one observes a monotonous increase of $m$ for higher
values of the average density. At the transition regime $m$
converges smoothly to zero if a finite value of the braking noise is
considered. This behaviour is not the result of a finite system
size, the data shown in Fig. \ref{fig_op} show no size
dependence anymore. Due to the smooth convergence of the
order-parameter it is difficult to determine a transition density, but
it is obvious that interactions between the cars occur already below
the density of maximum flow.

In the deterministic limit ($p=0$) the situation is
different. Here one observes a sharp transition at $\rho_c =
\frac{1}{v_{max}+1}$, the density of maximum flow.
%%%%%%%%%%%%%%%%%%%%%%%%%%%%%%% 
\section{Spatial Correlations}
%%%%%%%%%%%%%%%%%%%%%%%%%%%%%%%
A striking feature of a second-order phase transition is the diverging
length scale at criticality. In order to determine the length scale
for a given density, we measured the density-density correlation function
\begin{equation}
  G(r) = \frac {1}{L} \sum_{i=1}^{L} < n_i n_{i+r}> - \rho ^2 .
\end{equation} 
\begin{figure}[h]
  \centerline{\psfig{figure=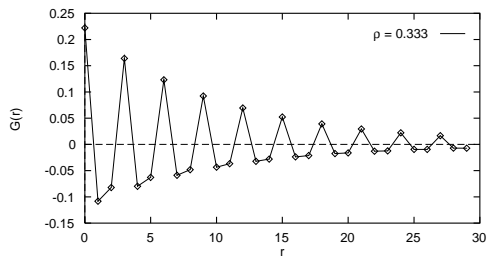,bbllx=120pt,bblly=220pt,bburx=500pt,bbury=430pt,height=3.5cm}}
  \caption{\protect{Correlation function in the presence of noise
      ($v_{max}=2, p = \frac{1}{128}$). The amplitude of the
      correlation function decays exponentially for 
      all values of $\rho$.}}
\label{fig_corr}
\end{figure}
Fig. \ref{fig_corr} shows the behaviour of the correlation function near the
transition density for $v_{max} = 2$ and $p = \frac{1}{128}$. The
correlation function is a damped oscillation with period
$v_{max}+1$. The amplitude is decaying exponentially for all values of
the average density if a finite value of $p$ is considered. 
In the deterministic limit one gets precisely at $\rho = \rho_c$  the
following behaviour of the correlation function:
\begin{equation}
  \label{korrdet}
   G(r) = \cases{\rho_c-\rho_c^2 & for\ $r \equiv 0$ mod$(v_{max}+1)$  \cr
-\rho_c^2  & else \cr}
\end{equation}
Due to the exponential decay one can determine the density dependence
of the  correlation length $\xi(\rho)$  if $p>0$. 

\begin{figure}[h]
  \centerline{\psfig{figure=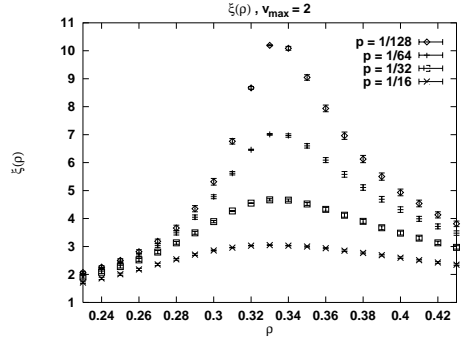,bbllx=90pt,bblly=230pt,bburx=570pt,bbury=590pt,height=4.5cm}}
  \caption{\protect{Density dependence of the correlation length in
      the vicinity of the transition.}}
\label{fig_xi}
\end{figure}
Fig. \ref{fig_xi} shows the behaviour of $\xi(\rho)$ for $v_{max}=2$ and
different values of $p$. The figure shows that the correlation length
has a maximum $\xi_{max}$ near the transition density, which is less
pronounced for higher values of the braking noise.

\begin{figure}[h]
  \centerline{\psfig{figure=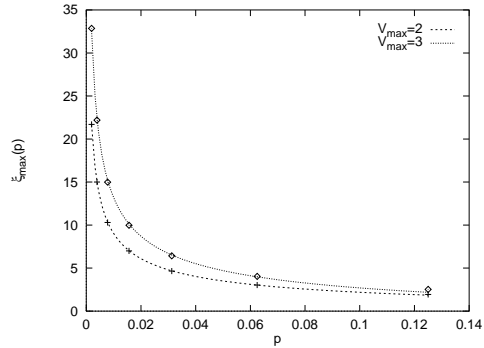,bbllx=50pt,bblly=50pt,bburx=550pt,bbury=400pt,height=4.5cm}}
  \caption{\protect{Noise dependence of $\xi_{max}$ for different
      maximum velocities. Independent of the maximum velocity, 
      $\xi_{max}(p) \sim 1/ \sqrt{p}$ holds in the limit $p
      \rightarrow 0$.}}
\label{fig_ximax}
\end{figure}
In Fig. \ref{fig_ximax} the noise dependence of $\xi_{max}(p)$ is
shown. The numerical data are in excellent agreement with  
\begin{equation}
  \label{ximaxasym}
  \xi_{max} \sim p^{-\frac{1}{2}} \qquad p \rightarrow 0.
\end{equation}

The divergence for $p=0$ is consistent with the behaviour of $G(r)$
at the transition density, where one observes a constant amplitude
of $G(r)$ and therefore an infinite correlation length. 
Finally it should be noted that the numerical results can be
confirmed analytically for $v_{max}=1$ \cite{duiproc}.

%%%%%%%%%%%%%%%%%%%        
\section{Summary}
%%%%%%%%%%%%%%%%%%%

We found evidence for the absence of criticality in the NaSch model
in the presence of noise. For finite $p$ the second order transition 
of the deterministic case is smeared out, similar to the situation of 
a equilibrium second order transition in a finite system. Here this effect
is caused by the presence of noise, $p>0$.

For small values of $p$ one finds an ordering transition close to
$\rho_c=1/(v_{max}+1)$. The ordering  is due the parallel
dynamics and can be found for all values of the maximum
velocity, including $v_{max}=1$. Larger values of the noise $p$ favour
the formation of jams 
and a tendency to phase separation occurs (see also
\cite{Luebeck,ZPR,Chowd}) if $v_{max}>1$. The demixing of free
flow and the jammed regime becomes more and more exact if larger values of
$v_{max}$ and $p$ are considered, because interactions between cars
become rare in the free flow regime of the system.  

The effect of fluctuations is twofold: First, one already finds blocked cars
for densities below $\rho_c$  and second, exact phase
separated states are not stable, e.g. one observes the spontaneous
formation of jams in the outflow region of a megajam.
  
Finally we discuss the deterministic limit $p=1$. Here we find
metastable states for densities below $\rho_c = 1/3$ if $v_{max}
>1$. These metastable states consist of moving cars with at least two
empty cells in front. They are metastable in the following sense: If a
car stops due to a pertubation of the system, all other cars have to
stop after  $O(L)$ lattice updates. These metastable states can only be
found for $v_{max}>1$, because for $v_{max}=1$ obviously the cars
cannot move. The lack of metastable states
for $0<p<1$ is consistent with our observation that {\em exact} phase
separation is absent for this cases.\\ 
\bigskip

{\bf Acknowledgement\\}

We thank Bernd Eisenbl\"atter and Michael Schreckenberg for their
collaboration in early stages of this work. 
%

%
% ---- Bibliography ----
%

\end{document}